\documentclass[%
 reprint,
 amsmath,amssymb,
 aps,
]{revtex4-1}

\usepackage[dvipdfmx]{graphicx}
\usepackage{here}
\usepackage{dcolumn}
\usepackage{bm}

\begin{document}

\preprint{APS/123-QED}

\title{$^{77}$Se-NMR Study under Pressure on 12\%-S Doped FeSe}

\author{Takanori Kuwayama$^1$, Kohei Matsuura$^2$, Yuta Mizukami$^2$, Shigeru Kasahara$^3$, Yuji Matsuda$^3$, Takasada Shibauchi$^2$, Yoshiya Uwatoko$^4$ and Naoki Fujiwara$^{1, \ast}$}
\affiliation{$^1$Graduate School of Human and Environmental Studies, Kyoto University, Yoshida-Nihonmatsu-cho, Sakyo-ku, Kyoto 606-8501, Japan\\
$^2$Graduate School of Frontier Sciences, University of Tokyo, 5-1-5 Kashiwanoha, Kashiwa, Chiba 277-8581, Japan\\
$^3$Division of Physics and Astronomy, Graduate School of Science, Kyoto University, Kitashirakawa Oiwake-cho, Sakyo-ku, Kyoto 606-8502, Japan\\
$^4$Institute for Solid State Physics, University of Tokyo, 5-1-5 Kashiwanoha, Kashiwa, Chiba 277-8581, Japan}

\date{\today}
\begin{abstract}
The 12\%-S doped FeSe system has a high $T_{\rm c}$ of $30~\mathrm{K}$ at a pressure of $3.0~\mathrm{GPa}$.
We have successfully investigated its microscopic properties for the first time via $^{77}$Se-NMR measurements under pressure.
The antiferromagnetic (AFM) fluctuations at the optimal pressure ($\sim3~\mathrm{GPa}$) exhibited unexpected suppression compared with the AFM fluctuations at ambient pressure, even though the optimal pressure is close to the phase boundary of the AFM phase induced at the high-pressure region.
In addition, we revealed that the SC phase at an applied field of $6.02~\mathrm{T}$ exhibited a remarkable double-dome structure in the pressure-temperature phase diagram, unlike the SC phase at zero field.

\end{abstract}
\pacs{Valid PACS appear here}
\maketitle
Recently, iron chalcogenides, so-called 11 systems, have received much attention because of their unique phase diagrams.
In particular, FeSe undergoes nematic and superconducting (SC) transitions at $90~\mathrm{K}$ and $9~\mathrm{K}$, respectively, without any magnetism at ambient pressure \cite{Baek2015}, while an antiferromagnetic (AFM) phase exists in most iron-based superconductors, such as undoped or low carrier doped 1111 and 122 systems \cite{Fernandes2014}.

The pressure-temperature $(P-T)$ phase diagram for FeSe is complicated as obtained from the resistivity measurements \cite{Sun2016a}: the nematic phase disappears at $1.5~\mathrm{GPa}$, and an AFM phase with a dome structure is induced in the $P$-$T$ phase diagram instead.
The AFM phase overlaps with the nematic phase at the boundary in the $P$-$T$ phase diagram.
The SC transition temperature ($T_{\rm c}$ ) of $9~\mathrm{K}$ at ambient pressure goes up to $37~\mathrm{K}$ at $6.0~\mathrm{GPa}$.
In this pressure-induced AFM phase, a stripe-type spin configuration with the nesting vector ($\pi$, 0) has been suggested from NMR measurements \cite{Wang2016}. The Fermi surfaces of FeSe are constructed by a hole pocket at the $\Gamma$ point and elliptical electron pockets at the M point.
More information about the Fermi surfaces at ambient pressure has been obtained from the angle-resolved photoemission spectroscopy (ARPES) \cite{Shimojima2014, Kasahara2014, Suzuki2015, Watson2015a, Watson2017, Kushnirenko2018, Coldea2018}.
Several experiments suggest orbital ordering under the nematic states, where the degeneracy between $d_{xz}$ and $d_{yz}$ orbitals is resolved \cite{Shimojima2014, Watson2015a}.
A theoretical investigation proposed a scenario that an inner hole-like pocket appears due to increasing pressure and would induce the AFM ordering with the $(\pi, 0)$ nesting vector \cite{Yamakawa2017a, Skornyakov2018}.
	\begin{figure}[bp]
	      \centering
	   \includegraphics[width=7.5cm]{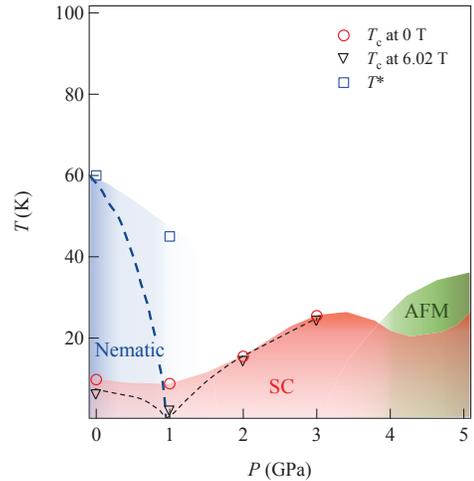}
	   \caption{(Color online) $P$-$T$ phase diagram for 12\%-S doped FeSe.
	   The phases in red, blue, and green represent the superconducting, short- or long-range nematic, and antiferromagnetic phases, respectively.
	   The blue dashed line represents the long-range nematic phase \cite{Matsuura2017}.
	   The black dashed line represents the SC phase at $6.02~\mathrm{T}$.
	   $T^{\ast}$ represents the temperature at which the FWHM shows the anomalous upturn.
	   Both $T_\mathrm{c}$ at $0~\mathrm{T}$ and $T_\mathrm{c}$ at $6.02~\mathrm{T}$ were determined from the AC susceptibility measurements.}
	   \label{figure:phase}
	\end{figure}

The phase diagram determined from the resistivity dramatically changes with sulfur (S) doping \cite{Matsuura2017}: the pressure-induced AFM phase with the dome structure moves to a higher pressure region as the doping level is increased, as shown in Fig.\ref{figure:phase}.
As a result, the nematic phase is separated from the AFM phase in the $P$-$T$ phase diagram.
Interestingly, $T_{\rm c}$ for $x = 0.12$ reaches a maximum ($\sim$30 K) at the intermediate pressure ($\sim$3 GPa) where both the nematic and AFM phases are absent \cite{Matsuura2017}.
Contrary to the $P$-$T$ phase diagram, no AFM phases are induced in the $x$-$T$ phase diagram at ambient pressure \cite{Watson2015b, Hosoi2016, Reiss2017}.
An additional hole pocket emerges, and the electron pockets become isotropic as the doping level is increased \cite{Coldea2016, Watson2015b, Reiss2017}.
To investigate the origin of the high $T_{\rm c}$, the 12\%-S doped sample is preferred to the pure sample because a high $T_{\rm c}$ of over $25~\mathrm{K}$ is attainable at low pressures ($\sim3~\mathrm{GPa}$), and is free from complex overlapping of nematic, SC, and AFM states.

To date, microscopic properties of the SC state with a high-$T_{\rm c}~(\sim25$-$30~\mathrm{K}$) have not been described. We have successfully investigated its microscopic properties for the first time via $^{77}$Se-NMR measurements under pressure.
Surprisingly, we found that the magnetic fluctuation at $3.0~\mathrm{GPa}$ is weaker than that at ambient pressure, although the AFM phase is induced in the high-pressure region.
To clarify the unexpected phenomenon, we carried out systematic NMR measurements under pressure and discovered an anomalous evolution of AFM fluctuations under pressure.

We performed $^{77}$Se-NMR measurements at $6.02~\mathrm{T}$ up to $3.0~\mathrm{GPa}$ on a 12\%-S doped single crystal with dimensions of approximately $1.0 \times 1.0 \times 0.5$ mm.
We used a NiCrAl pressure cell \cite{Fujiwara2007} and Daphne7373 as pressure mediation liquid.
The pressure was determined by Ruby fluorescence measurements \cite{Fujiwara2007}.
We placed the crystal in the pressure cell so that the FeSe plane was parallel to the applied field.

Fig.\ref{figure:phase} shows the $P$-$T$ phase diagram for 12\%-S doped FeSe.
$T^{\ast}$ plotted as blue circles in the phase diagram is anomaly in FWHM obtained from single Gaussian fits for $^{77}$Se-NMR spectra as mentioned below.
$T_{\rm c}$ was determined from AC susceptibility measurements using the tank circuit of an NMR probe at both $0~\mathrm{T}$ and $6.02~\mathrm{T}$.
The SC phase above $4.0~\mathrm{GPa}$ was extrapolated from the resistivity measurements \cite{Matsuura2017}.
At zero field, $T_{\rm c}$ was enhanced with increasing pressure and reached approximately $25~\mathrm{K}$ at $3.0~\mathrm{GPa}$.
$T_{\rm c}$ at $6.02~\mathrm{T}$ was almost the same as that at zero field except that at $1.0~\mathrm{GPa}$.
$T_{\rm c}$ at $1.0~\mathrm{GPa}$ is strongly suppressed, which results in double-dome structure in the $P$-$T$ phase diagram.

\begin{figure}[tbp]
	      \centering
	   \includegraphics[width=8.5cm]{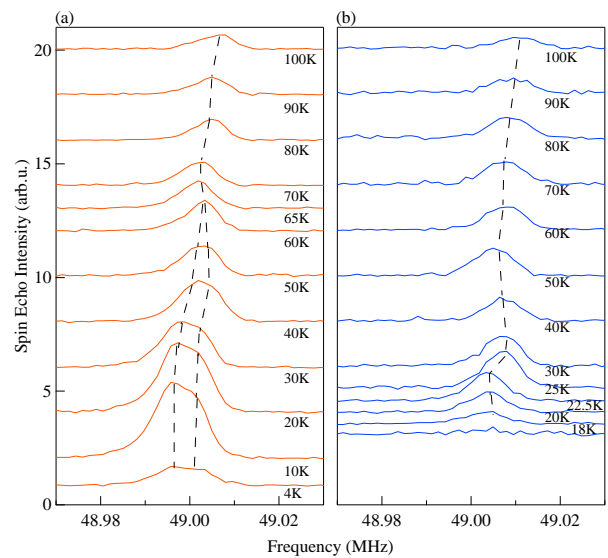}
	   \caption{(Color online) $T$ evolution of the $^{77}$Se-NMR spectra at (a) ambient pressure and (b) 3.0 GPa.
	   The black dashed lines represent the peak frequencies of each spectrum.
	   The signal intensity becomes extremely low below $T_{\mathrm c}(\sim24~\mathrm{K})$, and no signal was detected below $15~{\mathrm K}$ at 3.0 GPa.}
	   \label{figure:spectra}
	\end{figure}

We measured $^{77}$Se-NMR ($I = 1/2$, $\gamma / 2 \pi = 8.118~\mathrm{MHz/T}$) spectra on 12\%-S doped FeSe with a fixed field of $6.02~\mathrm{T}$ applied parallel to the crystallographic $a$ axis.
The $^{77}$Se-NMR spectra at ambient pressure and $3.0~\mathrm{GPa}$ are shown in Fig.\ref{figure:spectra}.
At ambient pressure, a single $^{77}$Se-NMR signal in a tetragonal state becomes a double-peak structure at approximately $60~\mathrm{K}$ in the nematic phase, which is in good agreement with the structural transition temperature obtained from resistivity measurements \cite{Matsuura2017}.
This double-peak structure disappears with increasing pressure (Fig.\ref{figure:spectra}b).
Below $T_{\rm c}$, the signal intensity becomes extremely small.
The signal of the pure sample $(x = 0)$ appears as two separated lines in the nematic phase \cite{Baek2015, Wang2016, Wiecki2018}.
According to the ARPES measurements and theoretical calculations\cite{Coldea2018, Reiss2017, Coldea2016}, the system becomes isotropic owing to S doping, which is consistent with the result that the splitting of the spectra becomes smaller than that for non-doped FeSe.

\begin{figure}[tbp]
	      \centering
	   \includegraphics[width=7.2cm]{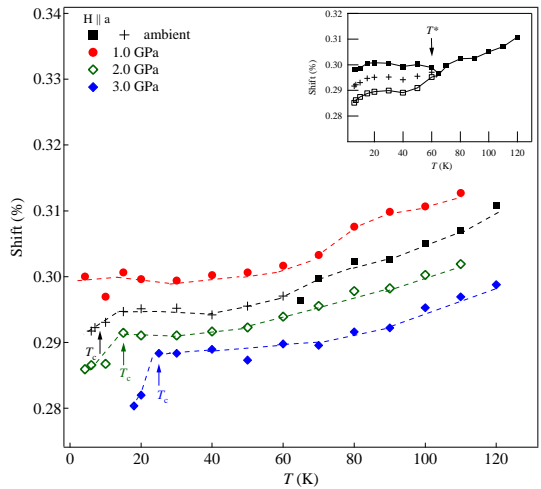}
	   \caption{(Color online) $T$ dependence of the $^{77}$Se-NMR shift at several pressures.
	   The black crosses represent the average of two lines in the nematic phase.
	   The inset shows the $T$ dependence of the shift at ambient pressure.
	   The split lines are obtained from two Gaussian fits.}
	   \label{figure:shift}
	\end{figure}

The $T$ dependence of $^{77}$Se shifts at several pressures are shown in Fig.\ref{figure:shift}.
The inset of Fig.\ref{figure:shift} shows the shift at ambient pressure.
The closed and open squares are determined from two Gaussian fits for the spectra.
The average of the double peaks at ambient pressure is plotted as black crosses in the main panel of Fig. \ref{figure:shift}.
The shift clearly drops just below $T_{\mathrm c}$ at $2.0$ and $3.0~\mathrm{GPa}$, which corresponds to $T_\mathrm{c}$ determined from the AC susceptibility measurements.
The shifts in Fig.\ref{figure:shift} qualitatively exhibit similar $T$ dependence above $T_{\mathrm c}$, and the quantitative difference comes from the $P$ dependence of the density of states (DOS).
In general, the DOS changes monotonically with increasing pressure owing to the change in the bandwidth.
In this case, however, the DOS is enhanced at $1.0~\mathrm{GPa}$, and then reduces with increasing pressure, indicating that some kind of anomaly in the Fermi surfaces occurs near $1.0~\mathrm{GPa}$.

\begin{figure}[tbp]
	      \centering
	   \includegraphics[width=7.2cm]{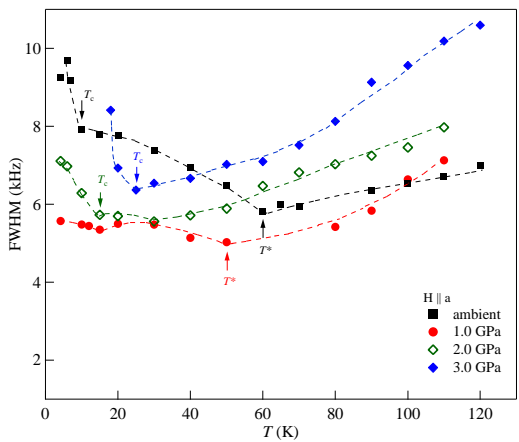}
	   \caption{(Color online) $T$ dependence of the full width at half maximum (FWHM) determined from a single Gaussian fit.
	   The dashed lines are guides for the eyes.}
	   \label{figure:FWHM}
	\end{figure}

Fig.\ref{figure:FWHM} shows the full width at half maximum (FWHM) obtained from a single Gaussian fit for the $^{77}$Se signal.
The FWHM tends to increase with increasing pressure, which may be due to the deterioration of hydrostaticity.
At ambient pressure, where the system undergoes the nematic phase below $60~\mathrm{K}$, the FWHM has two inflection points.
The first point at approximately $60~\mathrm{K}$ reflects the nematic transition; the second point at approximately $9~\mathrm{K}$ reflects the SC transition.
In addition, the $T$ dependence of the FWHM at ambient pressure shows a convex upward characteristic from $60~\mathrm{K}$ to $9~\mathrm{K}$, and a similar $T$ dependence was observed at $1.0~\mathrm{GPa}$ from $50$ to $15\mathrm{K}$.
According to the resistivity measurements \cite{Matsuura2017, Xiang2017}, the nematic quantum critical point may exist near $1.0~\mathrm{GPa}$.
Thus, the upturn seen in the FWHM at $1.0~\mathrm{GPa}$ may reflect short-range nematic order suggested in pure FeSe \cite{Wiecki2017a, Wang2017c}.
At $2.0~\mathrm{GPa}$, the FWHM takes almost constant values from $40~\mathrm{K}$ to $18~\mathrm{K}$, and shows a clear upturn at $T_{\mathrm c} \sim18~\mathrm{K}$.
At present, it is uncertain whether the short-range nematic order exists at $2.0~\mathrm{GPa}$.
At $3.0~\mathrm{GPa}$, the $T$ dependence of the FWHM shows a single upturn at $T_{\mathrm c}$, so that the nematic phase is completely absent.
The long- or short-range nematic transition temperature is plotted as $T^{\ast}$ in Fig.\ref{figure:phase}.

\begin{figure}[tbp]
	   \centering
	   \includegraphics[width=8.7cm]{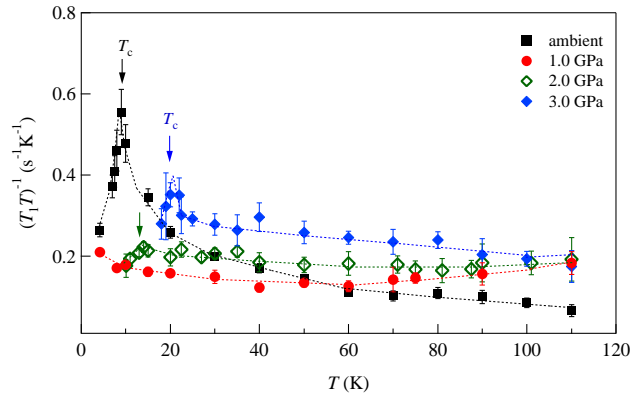}
	   \caption{(Color online) The relaxation rate $1/T_1$ divided by temperature $(T)$, $1/T_1T$, for $^{77}$Se measured at $6.02~\mathrm{T}$ up to $3.0~\mathrm{GPa}$.
	   The dashed lines are the guides for the eyes.}
	   \label{figure:T1T}
	\end{figure}

We measured the relaxation time $T_1$ with the inversion-recovery method.
$T_1$ at each $T$ for several pressures was determined from the nuclear magnetization $M_t$ using single-exponential function for $I = 1/2$: $1-M_t/M_0 = e^{-t/T_1}$, where $M_0$ is the equilibium nuclear magnetization.
The effect of local distribution due to 12\%-S doping was not observed.
The relaxation rate provides a measure of low-energy spin fluctuations.
When the wave vector $(q)$ dependence of the hyperfine interaction is neglected, $1/T_1T$ is expressed as
	\begin{equation}
	\frac{1}{T_1T} \propto \sum_q \frac{{\rm Im} \chi (q, \omega)}{\omega}
	\end{equation}
where $\omega$ and $\chi (q, \omega)$ represent the NMR frequency and dynamical spin susceptibility, respectively.
Fig.\ref{figure:T1T} shows $1/T_1T$ at several pressures.
The temperature where $1/T_1T$ has a peak is in good agreement with $T_{\rm c}$ determined from the AC susceptibility measurements.
Because the signal intensity becomes low below $T_{\rm c}$ as mentioned above (see Fig.\ref{figure:spectra}), we could not measure $T_1$ below $10~\mathrm{K}$ at $2.0~\mathrm{GPa}$ and $15~\mathrm{K}$ at $3.0~\mathrm{GPa}$.
At ambient pressure, $1/T_1T$ clearly shows Curie-Weiss-like behavior below $60~\mathrm{K}$ where the system undergoes the nematic transition.
The $T$ dependence at ambient pressure is similar to non-doped FeSe \cite{Imai2009, Wang2016, Wiecki2017a, Wiecki2018, Baek2015}.
On the other hand, in contrast to pure FeSe, the AFM fluctuations are strongly suppressed at $1.0~\mathrm{GPa}$ (Fig.\ref{figure:T1T}), whereas strong AFM fluctuations exist at ambient pressure.
In the high-pressure region, $1/T_1T$ is enhanced with increasing pressure, and at $3.0~\mathrm{GPa}$ the Curie-Weiss-like behavior revives below $30~\mathrm{K}$ (Fig.\ref{figure:T1T}).

In general, the AFM fluctuations determined from $1/T_1T$ strengthens toward an AFM phase.
However, the $P$ evolution of $1/T_1T$ in 12\%-S doped FeSe is non-monotonic.
The results of the shift in Fig.\ref{figure:shift} indicate that the $T$ dependence of $\chi (q=0)$ is qualitatively unchanged, which implies that the suppression of AFM fluctuations at $1$-$2~\mathrm{GPa}$ is intrinsic in this system.
This indicates that two different types of AFM fluctuations exist in the $P$-$T$ phase diagram.
Considering that the Fermi surfaces become isotropic\cite{Coldea2016, Watson2015b, Reiss2017, Skornyakov2018}, the Fermi surfaces may exhibit drastic variations in size and shape during pressurizing process.
Therefore, the topology of the Fermi surfaces may change as suggested in Ref. 18, and
the dominant nesting vector may change with increasing pressure.

According to recent inelastic neutron-scattering experiments for FeSe\cite{Wang2016b}, both stripe $(\pi,0)$ and N\'eel $(\pi, \pi )$ spin fluctuations were observed over a wide energy range.
In addition, the stripe-type spin configuration is suggested from NMR measurements for the pressure-induced AFM phase in FeSe\cite{Wang2016}.
Thus, the $(\pi, \pi )$ spin fluctuation may decrease and the $(\pi,0)$ spin fluctuation may increase with increasing pressure, which is a possible explanation for why the $P$ evolution of $1/T_1T$ in 12\%-S doped FeSe is non-monotonic.
To clarify this scenario, measurements with higher pressures and different S concentrations are needed.

In summary, we investigated the SC state with a high $T_{\rm c}$ of $25$-$30~\mathrm{K}$ for the first time.
We obtained the unexpected results; the AFM fluctuations at the optimal pressure are suppressed, although the optimal pressure is close to the AFM phase in the high-pressure region.
However, the AFM fluctuation at ambient pressure is predominant.
This implies that the dominant nesting vector may change with increasing pressure.

\section*{Acknowledgments}
The NMR work is supported by JSPS KAKENHI Grant Number JP18H01181 and a grant from Mitsubishi Foundation. We thank H. Kontani and P. Toulemonde for discussion.


\end{document}